\documentclass[aps,twocolumn,floatfix,prl,longbibliography]{revtex4-1} 

\usepackage{amsmath,mathtools,amssymb,latexsym, 
epic, epsfig, rotating, amsthm, todonotes }
\usepackage{titlesec}
\usepackage{color}
\usepackage{soul}
\usepackage{morefloats}
\newcommand{\bi}[1]{Fig.~\ref{fig:#1}}
\newcommand{\lr}[1]{\left\langle #1 \right\rangle}
\newcommand{\e}[1]{Eq.~(\ref{eq:#1})}
\newcommand{\be}{\begin{equation}} \newcommand{\ee}{\end{equation}}
\newcommand{\ba}{\begin{eqnarray}} \newcommand{\ea}{\end{eqnarray}}

\definecolor{darkgreen}{rgb}{0,0.6,0}
\definecolor{orange}{rgb}{0.99,0.257,0}

\usepackage{graphicx}
\usepackage{amsfonts, amsmath, amssymb, amsbsy}
\usepackage{siunitx}
\usepackage{setspace}
\usepackage{hyperref}
\usepackage{cleveref}
\crefname{appsec}{{}}{{}}

\AtBeginDocument{%
	\newwrite\bibnotes
	\def\bibnotesext{Notes.bib}
	\immediate\openout\bibnotes=\jobname\bibnotesext
	\immediate\write\bibnotes{@CONTROL{REVTEX41Control}}
	\immediate\write\bibnotes{@CONTROL{%
			apsrev41Control,author="48",editor="1",pages="1",title="0",year="1"}}
	\if@filesw
	\immediate\write\@auxout{\string\citation{apsrev41Control}}%
	\fi
}%

\allowdisplaybreaks[0]

\DeclareMathOperator{\erfc}{erfc}

\begin{document}

\title{
Fluctuation-dissipation relations for spiking neurons}

\author{Benjamin Lindner} 
\affiliation{Bernstein Center for Computational Neuroscience Berlin, Philippstr.\ 13, Haus 2, 10115 Berlin, Germany}
\affiliation{Physics Department of Humboldt University Berlin, Newtonstr.\ 15, 12489 Berlin, Germany}
\date{\today}

\begin{abstract}
Spontaneous fluctuations  and stimulus response are essential features of neural functioning but how they are connected is poorly understood. I derive fluctuation-dissipation relations (FDR) between the spontaneous spike and voltage correlations  and the firing rate susceptibility for i) the leaky integrate-and-fire (IF) model with white noise; ii) an  IF model with arbitrary voltage dependence, an adaptation current, and correlated noise. The FDRs can be used to derive correlation statistics or to infer the system's response  from observations of its spontaneous activity.
\end{abstract}
\maketitle

Small physical systems often display considerable fluctuations that can be characterized by correlation functions or power spectra. Fluctuation-dissipation relations connect the statistics of these spontaneous fluctuations of certain observables to their mean response to a time-dependent perturbation. Originally proposed for  equilibrium thermodynamic systems \cite{CalWel51,Kub66}, they have been extended to nonequilibrium setups with a steady state \cite{Aga72,HanTho82,ProJoa09,GomPet09}. Fluctuation-dissipation theorems can be used to infer the response properties from observations of purely spontaneous activity, to prove in a model-free way that a system operates outside thermodynamic equilibrium \cite{MarHud01,MizTar07}, or  to test whether a system obeys a Markovian description \cite{DinMar12,WilSok17}; for general reviews on applications of FDRs, see the comprehensive reviews \cite{MarPug08,Sei12}. 

Fluctuations are especially prominent in neural systems, specifically in the spike generation of neurons (nerve cells) in the brain, which is reflected in a long history of stochastic modeling in neuroscience \cite{Tuc89,GerKis14}. Neurons are notoriously noisy due to intrinsic sources of fluctuations (e.g. channel noise and unreliable synaptic transmission); in the recurrent networks of the cortex, the nonlinear interactions among many pulse-generating units lead to a strong chaotic variability (a network noise) even if single units follow a completely deterministic dynamics (i.e. the above mentioned channel noise, for instance, is neglected) and even if external (noisy) stimulation is absent. Most importantly, the response to external signals are of overarching importance for nerve cells, as it characterizes the transmission and processing of information, which is the main task of these cells. So, it is of vital importance to understand potential connections between the statistics of spontaneous activity and the response to a time-dependent perturbation in the case of spiking neurons.
 
Let us consider a paradigmatic stochastic model of computational neuroscience, the leaky integrate-and-fire model with white noise $\xi(t)$ and a time-dependent current signal $s(t)$:
\be
\frac{dv}{dt}=-v+\mu+s(t) +\sqrt{2D}\xi(t).
\label{eq:LIF-model}
\ee
The voltage across the nerve membrane, $v(t)$, upon reaching a threshold $v_T$, is reset to $v_R<v_T$ and, simultaneously, the time instant is registered as a spike time $t_i$.  The most important output of this model is the spike train, $x(t)=\sum \delta(t-t_i)$ (this is what is communicated to other cells). In \e{LIF-model} time and voltage are measured in multiples of the membrane time constant $\tau_m$ and the threshold-reset distance, respectively. The mean constant input $\mu$ and the intensity of the white noise $D$ are important parameters that determine the stochastic regime of the model \cite{VilLin09b}.

For the spontaneous activity ($s(t)\equiv 0$) the power spectrum of the spike train can be analytically calculated and expressed in terms of parabolic cylinder functions $\mathcal{D}_{a}(x)$ (see e.g. \cite{Lin02}):
\be
\label{eq:spec-LIF}
S_{xx}(\omega) = r_{0} \frac{
	|
	\mathcal{D}_{i\omega}(z_T
	)
	|^2
	- e^{\frac{z_R^2-z_T^2}{2}}
	|
	\mathcal{D}_{i\omega}(
	z_R)
	|^2
}{
	|
	\mathcal{D}_{i\omega}(
	z_T
	)
	- e^{\frac{z_R^2-z_T^2}{4}}
	\mathcal{D}_{i\omega}(
	z_R
	)|^2
}.
\ee
Here $z_{T/R}=(\mu-v_{T/R})\sqrt{D}$ and $r_0 = \lr{x(t)}$ is the stationary firing rate  \cite{note_Lindner_FDR_spiking_2022_1} with the angular brackets indicating an ensemble average.  

The response to a weak signal $s(t)$ is quantified by the time-dependent rate modulation $r(t)\approx r_0+ K_x * s(t)$, given in terms of a convolution with the linear response function $K_x(t)$ or in terms of the susceptibility $\chi_x(\omega)$ (the Fourier transform of $K_x$), which can be expressed by confluent hypergeometric functions \cite{BruCha01} or, equivalently, again in terms of parabolic cylinder functions \cite{LinLSG01}:
\be
\label{eq:susci-LIF}
\chi_{x}(\omega)\!=\! \frac{}{}
\frac{i r_{0}\omega/\sqrt{D}}{i \omega - 1}
\frac{
	\mathcal{D}_{i\omega - 1}
	(
	z_T
	)
	- e^{\frac{z_R^2-z_T^2}{4}}
	\mathcal{D}_{i\omega - 1}(
	z_R
	)
}{
	\mathcal{D}_{i\omega}
	(
	z_T
	)
	- e^{\frac{z_R^2-z_T^2}{4}}
	\mathcal{D}_{i\omega}
	(
	z_R
	)
}.
\ee
There is some structural similarity in the expressions for power spectrum and susceptibility - both are given in terms of ratios of differences of parabolic cylinder functions, but, apparently, it is not possible to express one in a simple way by the other.  So, even in this case, where we know the explicit solutions for the two characteristics of spontaneous fluctuations and of the response to a stimulus, it does not help us to connect them in a fluctuation-dissipation relation. The situation is similar (analytical expressions are known but cannot be related) for IF models with shot noise \cite{RicSwa10,DroLin17}, with dichotomous background noise \cite{DroLin17a}, or escape noise \cite{DegSch14}.

Here we connect the statistics of spontaneous spiking and the firing rate response to a weak signal by means of a simple calculation, which is markedly different to the typical derivation of the standard FDR \cite{note_Lindner_FDR_spiking_2022_2} and also to recent calculations for IF models in discrete time and embedded in networks \cite{CesAmp21}. The approach here builds on two ideas: i) the reset can be incorporated into the Langevin dynamics by means of the spike train (see e.g. \cite{LaiCho01} or \cite{MefBur04}), which permits to average this and related equations, leading by the Rice method to equations for spectral measures; ii) by means of the Furutsu-Novikov theorem \cite{Fur63,Nov65}, we can relate the noise-spike-train correlator to the exact linear response function. We first outline this calculation for the simple model in \e{LIF-model} and then treat the biophysically more realistic and dynamically richer exponential integrate-and-fire model endowed with correlated (colored) current noise. 

\emph{LIF model with white noise.--} Without signal current ($s(t)\equiv 0$), we can rewrite \e{LIF-model} as follows
\be
\frac{dv}{dt}=-v+\mu +\sqrt{2D}\xi(t) - (v_T-v_R)x(t),
\label{eq:LIF-model2}
\ee
where the last term formally imposes the fire-and-reset rule: the Delta-functions in $x(t)$ will push the voltage back from the threshold at $v_T$ to the reset point $v_R$. Having incorporated the reset rule into the equation, we can now take averages over a stationary ensemble. If we, for instance, directly average \e{LIF-model2}, we obtain
\be
\frac{d}{dt}\lr{v}=0=\mu-\lr{v}-(v_T-v_R) r_0.
\label{eq:direct-average}
\ee
which leads to $\lr{v}=\mu-(v_T-v_R) r_0$, a non-trivial relation between mean  membrane voltage and firing rate. 

Next, we take \e{LIF-model2} at time $t+\tau$, multiply with the spike train $x(t)$ at time $t$, and average. Using \e{direct-average} and expressing the derivative by one w.r.t.  $\tau$, we arrive at
\be
\frac{dC_{xv}}{d\tau} =-C_{xv}-(v_T-v_R) C_{xx}+\sqrt{2D} \lr{x(t)\xi(t+\tau)}.
\label{eq:LIF-DGL}
\ee  
Here we have introduced general correlation functions $C_{yz}(\tau)=\lr{y(t)z(t+\tau)}-\lr{y}\lr{z}$ for time series $y(t)$ and $z(t)$. One crucial insight is now that the last term in the above equation, the spike-train-white-noise correlation, is exactly proportional to the susceptibility with respect to a weak signal \emph{in the presence of the white background noise of intensity $D$}. In fact, as a consequence of the Furutsu-Novikov theorem this would hold true for any (in general, correlated) input noise as long as it is Gaussian. To see this more directly for our problem, imagine the (Gaussian) noise $\xi(t)= \sum \xi_n(t)$ being subdivided into $N$ independent Gaussian processes $\xi_n(t)$ with identical power spectra $S_{nn}=S_{\xi,\xi}/N$. The above correlation is then the sum of the single correlation functions  $\lr{x(t)\xi_n(t+\tau)}$, in which $\xi_n(t)$ represents a perfectly weak signal that is transmitted in the presence of a noise of intensity $D$ (all the other $N-1$ terms amount now to practically the entire noise process $\xi(t)$). In result, we obtain $\sqrt{2D} \lr{x(t)\xi(t+\tau)}=2D K_x(-\tau)$ or, in the Fourier domain, 
$\sqrt{2D} S_{x\xi}=S_{\xi\xi} \chi_x(\omega)$. Fourier transformation of \e{LIF-DGL} then yields the following fluctuation-dissipation relation for a stochastic LIF neuron with white noise
\be
\chi_x(\omega)=\frac{(v_T-v_R)S_{xx}(\omega)+(1+i\omega)S_{xv}(\omega)}{2D}
\label{eq:FDR-LIF}
\ee

\begin{figure}[h!]
	\includegraphics[width=0.4\textwidth]{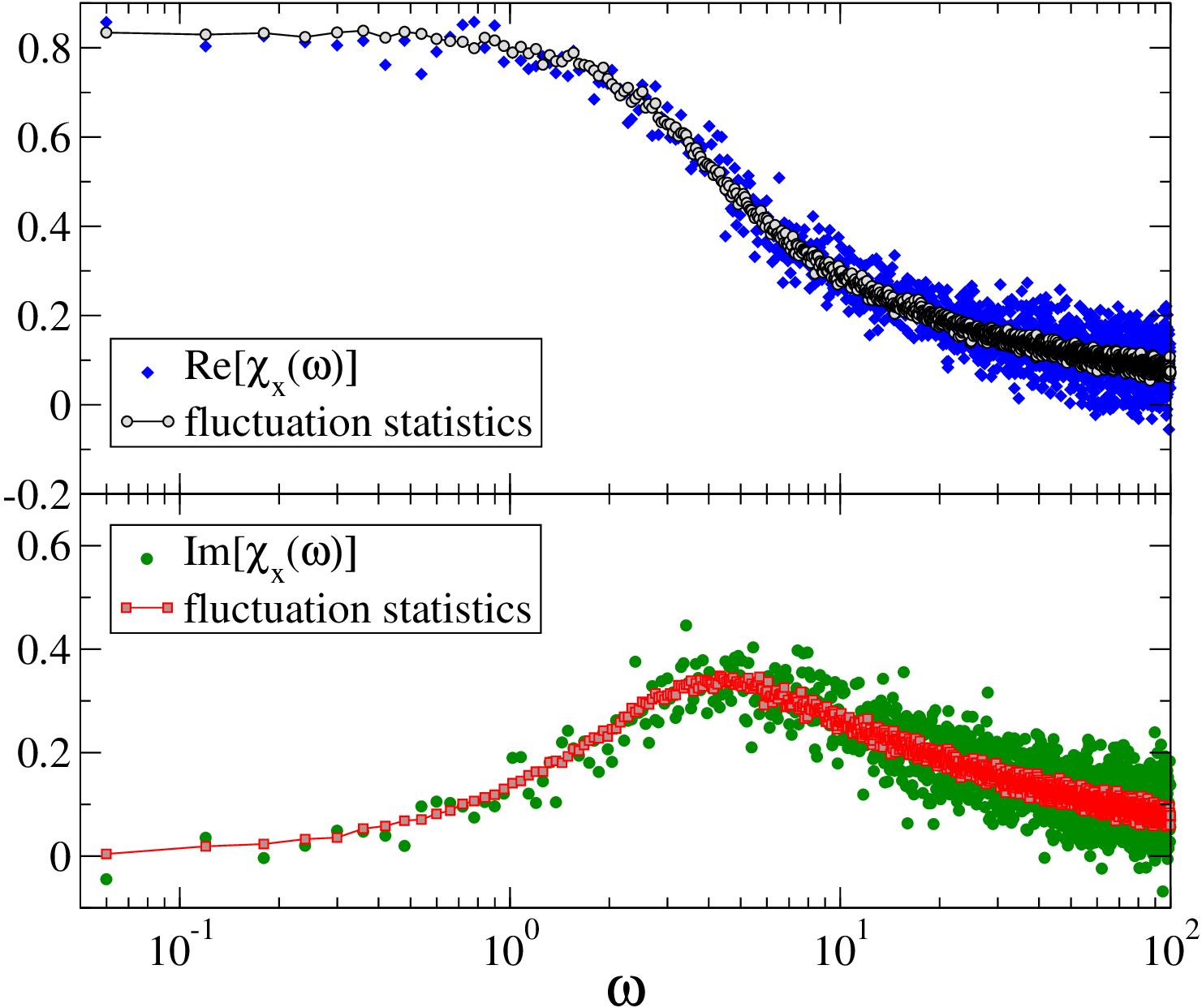}
	\caption{\textbf{Confirmation of the FDR for a white-noise driven leaky IF model.} Real  (top) and imaginary part (bottom) as functions of frequency for the left side (response properties) and the right side (spontaneous activity) of \e{FDR-LIF} for $\mu=0.8, D=0.1$ and a broadband stimulus (uniform power for $|\omega|<2\pi\cdot 100$) and a small variance of $\lr{s^2(t)}=0.1$. For both sets of simulations, $10^4$ trials, a time step of $\Delta t=10^{-4}$ and a time window of $T\approx 100$ were used (for real neurons with $\tau_m=10$ms, this would translate into a time window of 1s). 
	 	}
	\label{fig:LIF-FDR-confirm} 
\end{figure}

On the left hand side, we find the susceptibility of the firing rate with respect to a weak time-dependent signal, as can be, for instance, determined by a periodic stimulation $s(t)=\varepsilon \cos(\omega_s t)$ from the rate modulation $r(t)=\lr{x(t)}=r_0 +|\chi_x(\omega_s)|\cos(\omega_s t -\arg(\chi_x(\omega_s))$ (here $\arg(\cdot)$ is the complex argument)  or, equivalently, by a weak broadband stimulus as was used in \bi{LIF-FDR-confirm}.
On the right side of \e{FDR-LIF} are statistics of the spontaneous activity ($s(t)\equiv 0$): Besides the spike train power spectrum, the cross-spectrum between the subthreshold membrane voltage and the generated spikes emerges - this is the missing link between the spontaneous fluctuation statistics and the response statistics. For a selected parameter set, the relation is tested and confirmed in \bi{LIF-FDR-confirm}. 

Because we know most of the statistics by explicit expressions, we can use the relation above to determine the cross-spectrum between $v(t)$  and $x(t)$ analytically:
\be
S_{xv}(\omega)=\frac{2D \chi_x(\omega)-(v_T-v_R)S_{xx}(\omega)}{1+i\omega},
\label{eq:cross-LIF}
\ee  
which by virtue of \e{spec-LIF} and \e{susci-LIF} can be expressed by parabolic cylinder functions and is confirmed in \bi{LIF-cross} by numerical simulations.
\begin{figure}[h!]
	\includegraphics[width=0.4\textwidth]{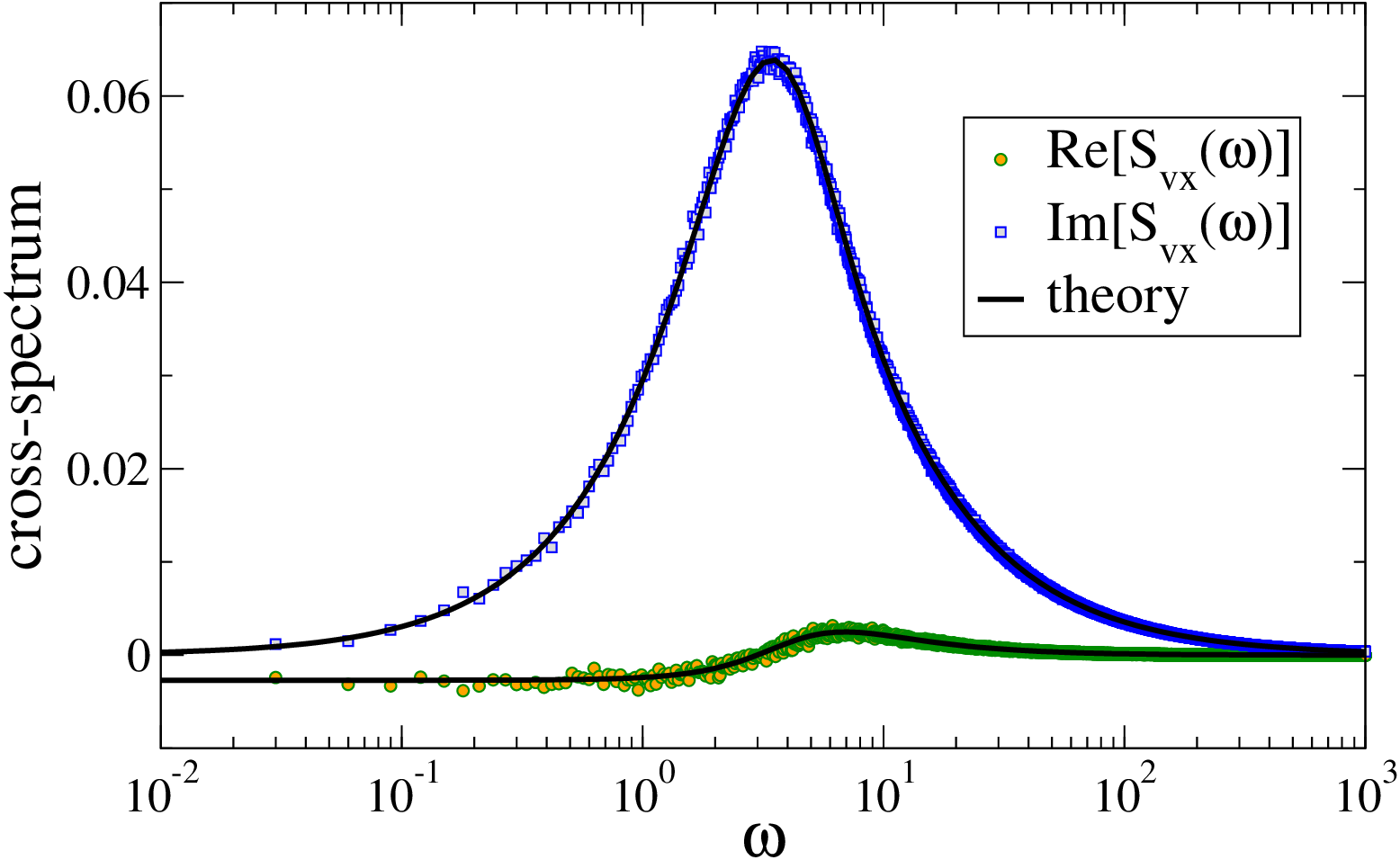}
	\caption{\textbf{Cross-spectrum between subthreshold voltage and spike train.}
	 	Parameters as in \bi{LIF-FDR-confirm}.}
	\label{fig:LIF-cross} 
\end{figure}

The form \e{cross-LIF} is instructive because it clearly shows the two sources of cross-correlation between membrane voltage and spike train. For once, there are the reset events occurring as steps at the spike times: The voltage contains so to speak the integrated spike train and is thus in part correlated to the spike train $x(t)$ as $x(t)$ is correlated to itself. Secondly, the correlation with the noise is shared between voltage and spike train and accounted for by the response function.

\emph{Exponential IF model with adaptation current and colored noise.--} We now turn to a more general and biophysically more realistic model, which, in its essential ingredients, has been justified on theoretical grounds \cite{FouHan03} but also extracted from data \cite{BadLef08}. As suggested by Brette and Gerstner \cite{BreGer05}, we include a spike-triggered adaptation current \cite{BenHer03}, and instead of white noise we allow for a Gaussian noise with arbitrary temporal correlations:
\ba
\frac{dv}{dt}&=&f(v) -a +\eta(t) - (v_T-v_R)x(t)+s(t),\\
\tau_a \frac{da}{dt}&=&-a + \Delta_a\tau_a x(t).
\ea
Here the white noise $\xi(t)$ has been replaced by a colored noise $\eta(t)$ with a prescribed power spectrum $S_{\eta\eta}(\omega)$. The simple linear leak term has been replaced with the function $f(v)=\mu-v+\Delta_v \exp((v-v_t)/\Delta_v$ (see \cite{FouHan03}). We note that the parameter $v_t<v_T$ sets a kind of soft threshold but we still keep a hard threshold at $v_T$ and a corresponding reset rule (this has been already incorporated above). 
%The threshold parameter $\Delta_v$ has interesting limits: for $\Delta_v\to 0$ we recover the leaky IF model; for $\Delta_v\to\infty$ the model approaches a quadratic IF model.
The variable $a(t)$ acts as an inhibitory current that pushes the voltage away from threshold. 
It evolves according to the slow dynamics given in the second equation (the ratio of its time constant to the membrane time constant is typically $\tau_a \gg 1$) but every spike generated by the model kicks the adaptation variable up by the amount $\Delta_a$, which implements the negative feedback that results in the spike-frequency adaptation seen in so many brain cells \cite{BenHer03}.

We can again use the same methods to derive equations for correlation functions and cross- and power spectra: Multiplication with $x(t)$ of the two eqs. above taken at time $t+\tau$, averaging, using expressions for the mean values, and, finally, expressing the noise-spike-train correlation function by the linear response susceptibility. This gives us a fluctuation-dissipation relation for the adapting exponential integrate-and-fire model: 
\be
\chi_x=\frac{ \left(v_T-v_R+\frac{\Delta \tau_a}{1+i\omega \tau_a}\right)S_{xx}+  i\omega S_{xv}-S_{xf(v)}}{S_{\eta\eta}}
\label{eq:FDR-EIF}
\ee
Again, on the left hand side we have exclusively the response to a weak stimulus, which  can be determined by means of a periodic or a broadband stimulus $s(t)$. On the right hand side, we find exclusively statistics of the spontaneous activity for $s(t)\equiv 0$. Several observations can be made: i) the adaptation dynamics enters only by modifying the prefactor of the spike train power spectrum in a frequency-specific manner; ii) instead of only the cross-spectrum of the subthreshold membrane voltage and the spike train (which still appears on the r.h.s.), we now also get the cross-spectrum of the spike train $x(t)$ and the subthreshold nonlinearity $f(v(t))$, which in some situations can be extracted from experiments \cite{BadLef08}.

\begin{figure}[h!]
\includegraphics[width=0.4\textwidth]{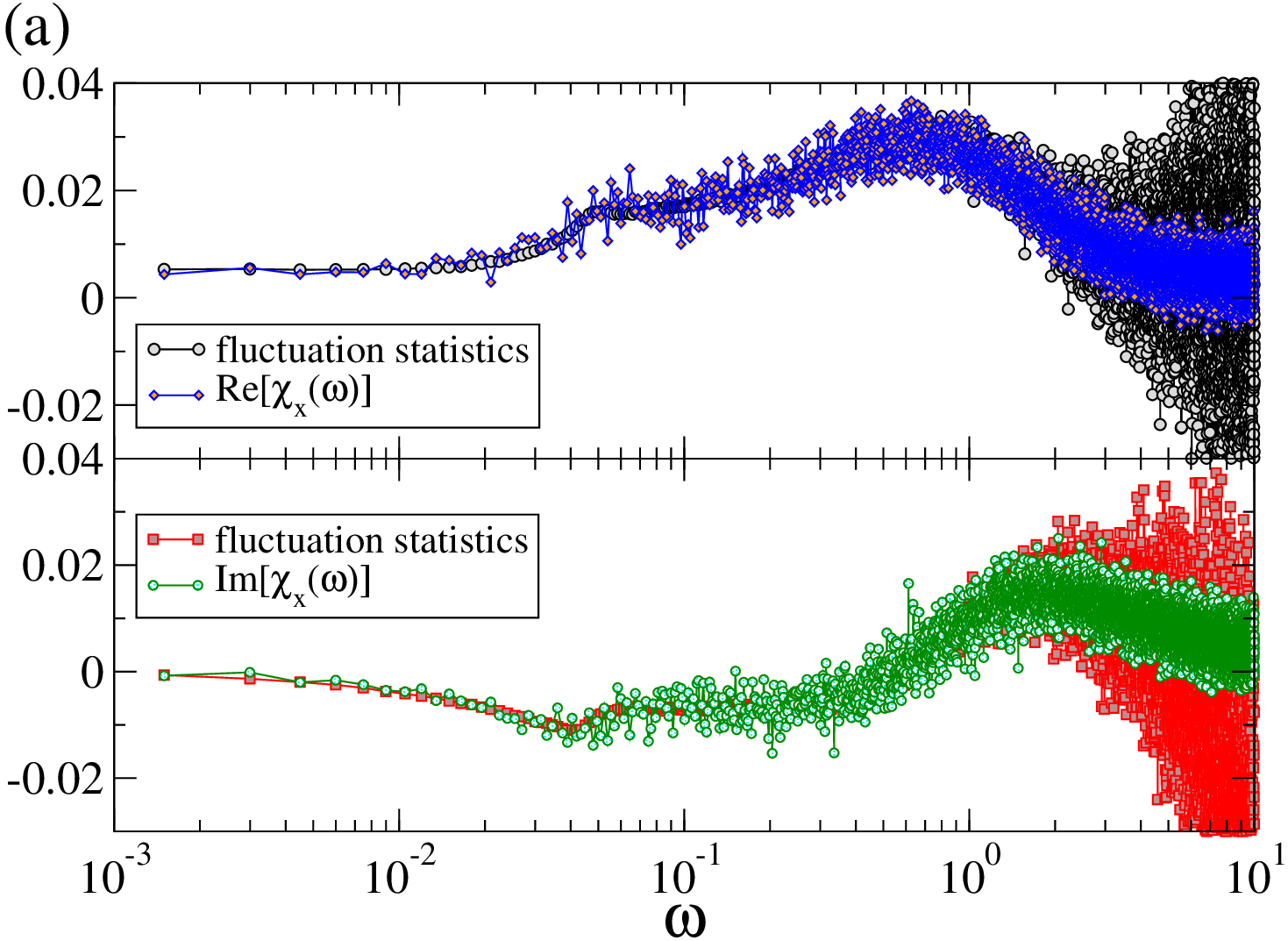}
	\includegraphics[width=0.41\textwidth]{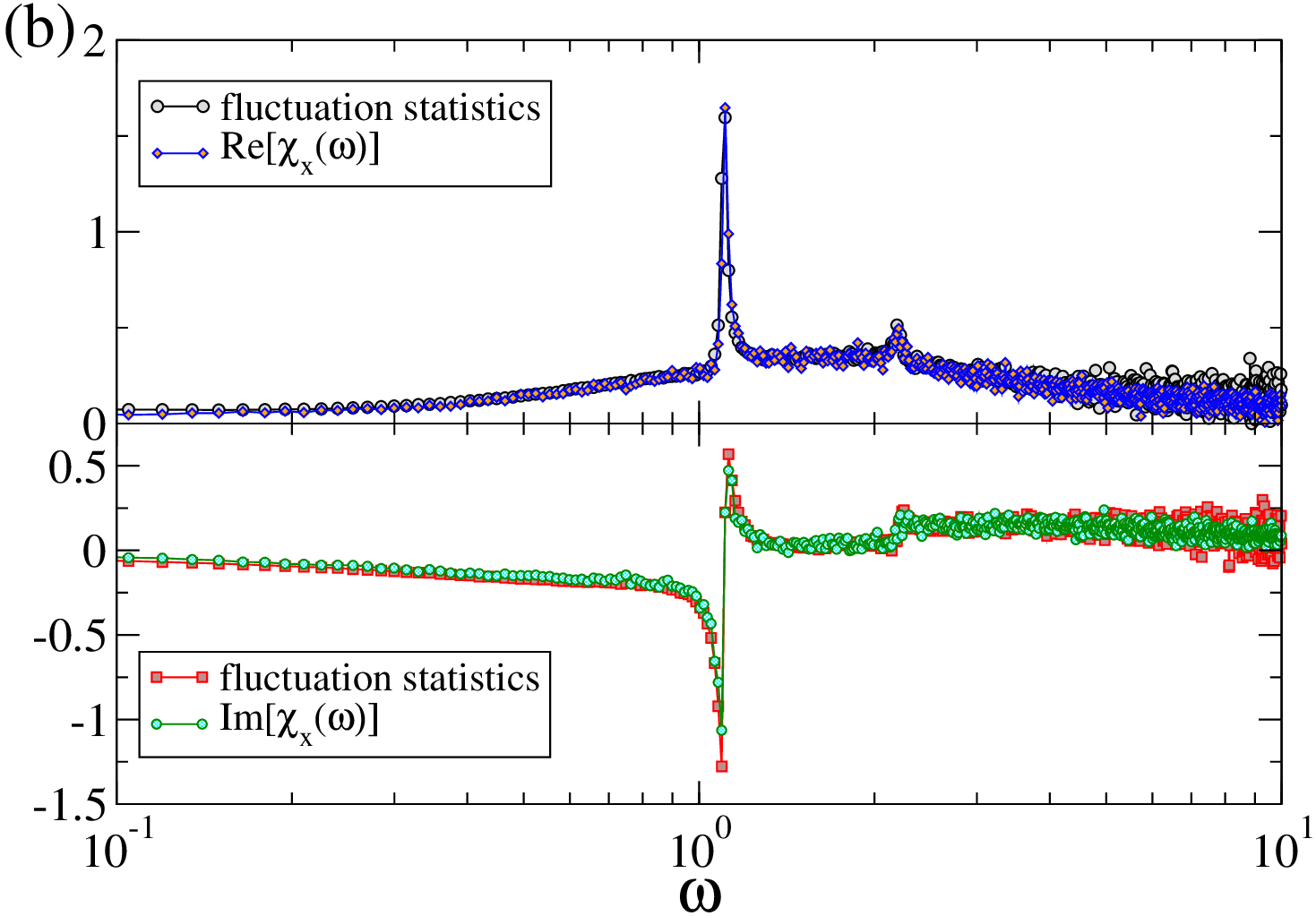}
	\caption{\textbf{Confirmation of the FDR \e{FDR-EIF} for an exponential IF model with a (colored) Ornstein-Uhlenbeck noise and a spike-triggered adaptation current.} Real  (top) and imaginary part (bottom) as functions of frequency for the left side (response property $\chi(\omega)$) and the right side (fluctuation statistics) of \e{FDR-EIF} for $\mu=0.8, \tau_{OUP}=10,\sigma^2=0.5, \tau_a=100, \lr{s^2(t)}=0.2$ in (a) and $\mu=4, \tau_{OUP}=1,\sigma^2=0.1, \tau_a=10, \lr{s^2(t)}=0.1$ in (b). In both plots $v_t=1, \Delta_v=0.2,\Delta_a=2, v_R=0, v_T=1$. For the determination of the susceptibility a broadband stimulus (uniform power for $|\omega|<2\pi\cdot 100$) was used. For all simulations, $10^4$ trials, a time step of $\Delta t=10^{-3}$ (in a) and $\Delta t=10^{-4}$ (in b) were used. 
	 	}
	\label{fig:EIF-FDR-confirm} 
\end{figure}
As examples we pick two cases in \bi{EIF-FDR-confirm}, which both confirm the FDR for the adapting neuron with colored noise. For both cases it is difficult to calculate analytically any of the statistics shown or used here.

 In \bi{EIF-FDR-confirm}a the constant input current $\mu$ is so weak that the model is still in the excitable regime (comparable to the LIF model) but we add a slow adaptation current and low-pass filtered noise (Ornstein-Uhlenbeck noise with an exponential correlation function and  a correlation time that is ten times the membrane time constant). In particular, the adaptation current leads to a high-pass shape of the susceptibility \cite{Sch13}. Remarkably, the numerical fluctuations of the susceptibility determined by stimulation or from the spontaneous statistics via \e{FDR-EIF} behave very differently: at small and up to intermediate frequencies ($\omega<1$) the r.h.s. of \e{FDR-EIF} provides a very reliable estimate of the susceptibility while the estimate obtained by broadband stimulation is more noisy. In the high-frequency limit, it is the other way around: the broadband stimulus yields a more reliable estimate of the response while the estimate from the spontaneous statistics becomes really noisy. Hence, if we want to know the response at specific frequencies, it will depend on the frequency whether we can benefit from \e{FDR-EIF} or whether we should resort to the simple stimulation method to get the susceptibility.

In \bi{EIF-FDR-confirm}b, the input current is increased, the Ornstein-Uhlenbeck noise is reduced in standard deviation and correlation time (now equal to the membrane time constant) such that the susceptibility displays pronounced resonances at the firing rate (i.e. at $\omega=2\pi r_0$). Also in this dynamically very different regime the relation between the fluctuation statistics and the response, \e{FDR-EIF}, is excellently confirmed. 

One potential problem with verifying and/or exploiting the relationship \e{FDR-EIF} for real neurons is that although the membrane voltage and the spike train might be accessible, the same is not necessarily true for the subthreshold nonlinearity $f(v)$. For non-adapting neurons, this can be determined by  \emph{in vitro} experiments \cite{BadLef08}, however, such experiments might not be possible or hampered by adaptation phenomena. Comparison of fluctuation and response may help to infer appropriate values of $v_t$ and $\Delta_v$, which would be one potential use for the FDR. 

\emph{Conclusions.--}  The relation between spontaneous fluctuations and the response to external perturbations have been worked out for an important class of spiking neuron models of the integrate-and-fire type. However, there is still work to be done: In the same framework, it is easily possible to derive relations for the susceptibility of the  subthreshold membrane voltage and its spontaneous power spectrum. The simple method introduced here can be applied to all types of neuron model with spike-associated reset such as the two-dimensional Izhikevich model \cite{Izh07} or the generalized IF model \cite{RicBru03} if the voltage dynamics is driven by Gaussian noise. 
%Not as straightforward is to take into account an absolute refractory period in such relations.  

We note that an entirely independent set of fluctuation-dissipation relations can be derived by the more common approach  to nonequilibrium thermodynamic systems with a steady state  going back to Agarwal \cite{Aga72,HanTho82} and discussed also more recently in the literature \cite{ProJoa09,DinMar12,WilSok17}. This approach 
 will lead to relations in terms of the correlation function of the conjugated variable that is a highly nonlinear function of the membrane voltage steady state distribution. Finally, there is also work on fluctuation-dissipation-relations for integrate-and-fire models in discretized time by Cessac et al. \cite{CesAmp21}, the relation to results here has to be clarified.
  
The division into subthreshold voltage and spike train used in  the derived relations may appear  somewhat conceived and as an artifact of the integrate-and-fire framework. However, it is meaningful because the spike train represents the important signal that is communicated to other neurons. In this context, it will be also useful to generalize the analysis to  conductance-based neuron models of the Hodgkin-Huxley type. As the above mentioned multidimensional IF models approximate conductance-based models in many situations surprisingly well, I expect that the fluctuation-dissipation relations found here may hold true (at least approximately) also for these more detailed models of neural firing.

In the context of neural information transmission, it is worth mentioning that both susceptibility and spike-train power spectra appear in the coherence function which provides a frequency-resolved measure of signal transfer \cite{Lin16}. Also in stochastic mean-field theories of neural networks, both power spectra as well as susceptibilities play important roles (see e.g. \cite{Bru00,KnoLin22}). For these problems, fluctuation-dissipation relations for spiking neurons may be used to simplify, reformulate, and better understand the statistics of interest.

%\bibliographystyle{spphys}
%\bibliographystyle{apsrev}       % APS-like style for physics
%\bibliography{ALL_22_03_19}   % name your BibTeX data base

\begin{thebibliography}{39}
\bibitem[{not({\natexlab{b}})}]{note_Lindner_FDR_spiking_2022_2}
\bibinfo{note}{In the standard approach, the response to a step perturbation is
  expressed by the linear response on the one hand and by a Taylor expansion of
  the steady state under constant perturbation on the other hand}.
\end{thebibliography}

\end{document}